\documentclass[runningheads]{article}

\usepackage[latin2]{inputenc}
\usepackage[english]{babel}
\usepackage{amsfonts,amsmath,amssymb}
\usepackage{eucal}
\usepackage{xcolor}
\usepackage{graphicx,epic,eepic}
\usepackage[scriptsize]{subfigure}
\usepackage{booktabs}
\renewcommand{\cal}{\mathcal}%\mathrsfs
\renewcommand{\mathfrak}{\textfrak}

\newcommand{\bdim}{\begin{proof}}\newcommand{\edim}{\end{proof}}%
\newcommand{\beq}{\begin{equation}}\newcommand{\eeq}{\end{equation}}%
\newcommand{\textv}[1]{\textquotedblleft{#1}\textquotedblright}
\newcommand{\ta}{\left(}%
\newcommand{\qa}{\left[}%
\newcommand{\ga}{\left\{}%
\newcommand{\tc}{\right)}%
\newcommand{\qc}{\right]}%
\newcommand{\gc}{\right\}}%
\newcommand{\tonde}[1]{\ta{#1}\tc}
\newcommand{\quadr}[1]{\qa{#1}\qc}
\newcommand{\graff}[1]{\ga{#1}\gc}

\newcommand{\bgraff}[1]{\big\{{#1}\big\}}
    \newcommand{\appl}{\rightarrow}

    \renewcommand{\iff}{\Leftrightarrow}
    \newcommand{\wh}[1]{\widehat{{#1}}}%
    \newcommand{\wt}[1]{\widetilde{{#1}}}%
    \newcommand{\ol}[1]{\overline{{#1}}}%
    \renewcommand{\epsilon}{\varepsilon}
    \renewcommand{\phi}{\varphi}%
    
    \newcommand{\W}{\Omega}
    \newcommand{\pog}[2]{{}^{\phantom{O}}_{#1}p_{#2}^O}%
    \newcommand{\psg}[2]{{}^{\phantom{S}}_{#1}p_{#2}^S}%
    \newcommand{\afig}[2]{a_{\,\, \text{\raisebox{-0.8ex}{\scriptsize\begin{tabular}{r|l}
              \cline{1-1}
             \hspace{-0.4em}  \raisebox{-0.3ex}{${#1}$} \!\!& \hspace{-0.3em}  \raisebox{-0.3ex}{${#2}$}
            \end{tabular} \hspace{-0.85em}}} }}

    \newcommand{\sfig}[2]{s_{\,\, \text{\raisebox{-0.8ex}{\scriptsize\begin{tabular}{r|l}
              \cline{1-1}
             \hspace{-.4em}  \raisebox{-0.3ex}{${#1}$} \!\!& \hspace{-.3em}  \raisebox{-0.3ex}{${#2}$}
            \end{tabular} \hspace{-0.85em}}} }}

    \newcommand{\gao}{\textsc{g.a.o.}}%
    \newcommand{\gaox}{guaranteed annuity options }%
    \DeclareMathOperator{\de}{d\!}%
    \newcommand{\E}{\mathbb{E}}

    \newcommand{\F}{\cal{F}}

    \renewcommand{\P}{\mathrm{P}}
    \renewcommand{\geq}{\geqslant}
    \renewcommand{\leq}{\leqslant}
    \newcommand{\deq}{\mspace{+4.8mu}\raisebox{.29pt}{\textnormal{:}}\mspace{-5mu}=}
    
    \newcommand{\per}{\times}%
    
\title{A policyholder's utility indifference valuation model for the guaranteed annuity option}

\author{ Matheus R Grasselli\footnotemark[1] \  \and Sebastiano Silla 
\footnotemark[2]}

\begin{document}
%\mainmatter
%\pagestyle{myheadings} \setcounter{page}{1}

\maketitle

\begin{abstract}
Insurance companies often include very long-term guarantees in participating life insurance products, which can turn out to be very valuable. Under a guaranteed annuity options (\gao), the insurer guarantees to convert a policyholder's accumulated funds to a life annuity at a fixed rated when the policy matures. Both financial and actuarial approaches have been used to valuate of such options. In the present work, we present an indifference valuation model for the guaranteed annuity option. We are interested in the additional lump sum that the policyholder is willing to pay in order to have the option to convert the accumulated funds into a lifelong annuity at a guaranteed rate.\\

\smallskip

\scriptsize
\noindent {\bf J.E.L. classification.} \textsc{D91; G11; J26.}

\smallskip

\normalsize \end{abstract}

\renewcommand{\thefootnote}{\fnsymbol{footnote}} \footnotetext[1]{%
McMaster University, Canada. E-mail: grasselli@math.mcmaster.ca.} 
\footnotetext[2]{%
Polytechnic University of Marche, Italy. E-mail: s.silla@univpm.it }

%****************************************************************************
%****************************************************************************

%===================================================
\section{Introduction and literature review}
%===================================================

Insurance companies often include very long-term guarantees in participating life insurance products, which can turn out to be very valuable. Guaranteed annuity options (\gao) are options available to holders of certain policies that are common in U.S. tax-sheltered plans and U.K. retirement savings. Under these options, the insurer guarantees to convert a policyholder's accumulated funds to a life annuity at a fixed rated when the policy matures. Comprehensive introductions to the design of such options are offered by O'Brien \cite{OBrien_ART_GAOs}, Boyle \& Hardy \cite{BoyleHardy_ART_GuaranAnnOpts} \cite{BoyleHardy_WP_GuarnAnnOpts}, Hardy \cite{Hardy_BOOK_InvestmentGuarantees} and Milevsky \cite{Milevsky_LIB}. For concreteness, we will focus on the analysis of a particular type of policy, but the framework we use can be readily extended to more general products in this class.

\subsection{The design of the policy}

We analyze a standard contract designed as follows: at time $t_0=0$ the policyholder agrees to pay a continuous premium at a rate $P$ for an insurance policy maturing at $T$. The premium is deemed to be invested in a money market account with continuously compounded interest rate $r$, and the policyholder receives the corresponding accumulated funds $A$ at time $T$. We are interested in the additional lump sum $L_0$ that the policyholder is willing to pay at time $t_0$ in order to have the option to convert the accumulated funds $A$ into a lifelong annuity at a guaranteed rate $h$.

Between time $t_0$ and time $T$, the liabilities associated with such guaranteed annuity options  are related to changes in economic conditions and mortality patterns. A rational policyholder will only exercise the option at time $T$ if it is preferable to the annuity rates prevailing in the market at that time. As remarked by Milevsky and Promislow \cite{Milevsky_Promislow_2001}, the company has essentially granted the policyholder an option on two underlying stochastic variables: future interest rates and future mortality rates.

\subsection{Literature review}

The nature of guaranteed annuity options was firstly presented in Bolton et al. \cite{Bolton_ART_ReservingForAnnuityGuarantees} and O'Brien \cite{OBrien_ART_GAOs}. The liabilities under \gaox represent an important factor that can influence the solvency of insurance companies. In a stochastic framework, a first pioneering  approach was proposed by Milevski and Posner \cite{Milevsky_Posner_2001} and  Milevsky and Promislow \cite{Milevsky_Promislow_2001}. The literature concerning the valuation of \gaox in life insurance contracts has  grown and developed in several directions. Both financial and actuarial approaches handle implicit (\textv{embedded}) options: while the formers are concerned with risk-neutral valuation and fair pricing, the others focus on shortfall risk under an objective real-world probability measure. The interaction between these two ways was analyzed by Gatzert \& King \cite{GatzertKing_ART_AnalysisParticipatingLifeInsuranceContracts}.
The seminal approach of Milevsky \& Promislow \cite{Milevsky_Promislow_2001} considered the risk arising both from interest rates and  hazard rates. In this context, the force of mortality is viewed as a forward rate random variable, whose  expectation is the force of mortality in the classical sense. On the same line, the framework proposed by Dahl \cite{Dahl_ART_StochasticMortality} described the mortality intensity by an affine diffusion process. Ballotta \& Haberman \cite{Ballotta_Haberman_ART_ValuationGAOs} \cite{Ballotta_Haberman_ART_} analyzed the behavior of pension contracts with guaranteed annuity options when the mortality risk is included via a stochastic component governed by an Ornstein-Uhlenbeck process. Then, Biffis \& Millossovich \cite{Biffis_Millossovich_ART} proposes a general framework that examines some of the previous contributions.  For an overview on stochastic mortality, longevity risk  and guaranteed benefits, see also Cairns et al. \cite{CairnsBlakeDowd_ART_ATwoFactorModel} \cite{CairnsBlakeDowd_ART_PricingDeath}, Pitacco \cite{Pitacco_ART_SurvivalModels} \cite{Pitacco___ART_LongevityRiskInLivingBenefits} and Schrager \cite{Schrager_ART_AffineStochasticMortality}.
Finally, a different approach, based on the annuity price, was offered by Wilkie \cite{Wilkie_ART_ReservingPricingHedgingGAOs} and Pelsser \cite{Pelsser_ART_PricingAndHedgingGAOS} \cite{Pelsser_PRT_PricingHedgingGuaranteedAnnuityOptions}. In particular, Pelsser introduced a martingale approach in order to construct a replicating portfolio of vanilla swaptions. We also mention the related contributions of Bacinello \cite{Bacinello_ART_FairValuationLifeInsuranceContractsWithEmbeddedOptions},  Olivieri \& Pitacco \cite{OlivieriPitacco_ART_RenditeVitalizeLongevityRisk},  \cite{OlivieriPitacco_ART_AnnuitisationGuaranteeAndUncertainty}, \cite{OlivieriPitacco_ART_ForecastingMortality} and Pitacco \cite{MaroccoPitacco_ART_LongevityRisk}, Olivieri \cite{Olivieri_ART_UncertaintyInMortalityProgections}.

\subsection{Objective of the paper}

The present paper considers, for the first time to our knowledge, an indifference model to value guaranteed annuity options. The indifference model proposed here can capture at once the incompleteness characterizing the insurance market and the theory of the optimal asset allocation in life annuities toward the end of the life cycle.

The \textit{priciple of equivalent utility} is built around the investor's attitude toward the risk. Approaches based on this paradigm are now  common in financial literature concerning incomplete markets. In a dynamic setting the \textit{indifference pricing} methodology was initially proposed by Hodges and Neuberger \cite{HodgesNeuberger}, who introduced the concept of \textit{reservation price}. For an overview, we address the reader  to the following contributions and to the related bibliography: Carmona \cite{Carmona_BOOK_IndiffPricing}, Musiela and Zariphopoulou \cite{MusielaZariphopoulou_ART_AnExampleOfIndifferencePrices}, Zariphopoulou \cite{Zariphopoulou_BOOK_StochasticControlMethods}. Recently Young and Zariphopoulou \cite{Young_Zari_2002} and Young \cite{Young03}, applied the principle of equivalent utility to dynamic insurance risk.

Our argument is inspired by the theory on the optimal asset allocation in life annuities toward the end of the life cycle. For instance, we refer to Milewsky \cite{Milevsky98_ART_OptimalAssetAllocation}, \cite{Milevsky2001}, Milewsky \& Young \cite{MilevskyYoung2002_ART_OptimalAssetAllocation}, \cite{MilevskyYoung_2003_ART_AnnuitizationAssetAllocation}, \cite{MilevskyYoung2003_PRT_OptimalAnnuityPurchasing}, \cite{MilevskyYoung_2007_ART_AnnuitizationAssetAllocation}, Milevsky et al. \cite{MilevskyMooreYoung2006_ART_AssAllocAnnuityPurchaseAndFinancialRuin} and Blake et al. \cite{Blake_Cairns_Dowd_ART_AnnuitizationOptions}.
The model is developed in two stages. First we compare two strategies at time $T$, when the policyholder is asked to decide whether or not she wants to exercise the guaranteed annuity option. Next, we go back to $t_0$ and compare the expected utility arising from a policy with the guaranteed annuity option against a policy where no implicit options are included.

Assuming a utility of consumption with constant relative risk aversion and constant interest rates, we find that the decision to exercise the option at time $T$ and the decision to purchase a policy embedding a guaranteed annuity option reduce to compare  the guaranteed rate $h$ and the interest rate $r$. It turns out that the indifference valuation is based on two quantities: the actual value of the guaranteed continuous life annuity, discounted by the implicit guaranteed rate, and the actual value of a perpetuity discounted by the market interest rate.

\subsection{Organization of the paper}

The remainder of the paper is organized as follows. Section \ref{sec:ValuatingGAO} describes the financial and actuarial setting where the model is defined. We characterize the optimal exercise when the policy matures and the strategies at the initial time, when the policy is purchased. The end of this section gives the definition and the explicit formula for the indifference valuation of the guaranteed annuity option. Section \ref{sec:Numerics_Insights} show numerical examples for  the equivalent valuation depending on  different scenarios for the interest rate. It also present a discrete-time version for the numerical simulations.

%=================================================
 \section{The model}
        \label{sec:ValuatingGAO}
%=================================================

\subsection{The financial market}

%---------------------------------------------------------------
We assume a policyholder who invests dynamically in a market consisting of a risky asset with price given by
        \[dS_t = \mu S_t dt + \sigma S_t dW_t\]
with initial condition $S_{0}>0$, where $W_t$ is a standard Brownian motion on a filtered probability space
$(\W,\,{\cal F},{\cal F}_t,P)$ satisfying the usual conditions of completeness and right-continuity, and $\mu$ and $\sigma>0$ are
constants. Furthermore, the policyholder can invest in a risk--free bank account described by
        \[dB_t = rB_t dt\]
with initial condition $B_{0}=1$, where $r$ is a constant representing the continuously compounded interest rate.

The policyholder is assumed to consume at a instantaneous rate $c_t\geq 0$ per year, self-financing her position using the market gains she is able to realize. To this end,  assume the agent is initially endowed by a positive wealth $x$ and the process $X_t$ will denote the wealth process for $t\geq 0$. At each time $t\geq 0$, the policyholder chooses dynamically the amount $\pi_t$ to invest in the risky asset and, consequently, the amount $X_t - \pi_t$ to be invested in the  risk free asset. The processes $c_t$ and $\pi_t$ need to satisfy some admissibility conditions, which we specify in the  next sections.

The assumed financial market follows the lines of Merton \cite{Merton69}, \cite{Merton71}, \cite{Merton90} and can be generalized, at cost of less analytical tractability, following the contributions provided, for example, by Trigeorgis \cite{Trigeorgis_ART_TheNatureOptionInteraction}, Kim and Omberg \cite{KimOmberg_ART_DynamicNonmyopic}, Koo \cite{Koo_ART_ConsumptionAndPortfolioSelection}, S{\o}rensen \cite{Sorensen_ART_DynamicAssetAllocation} and Wachter \cite{Wachter_ART_PortfolioConsumptionDecision}. For example, in Grasselli \& Silla \cite{GrasselliSilla_BuyerGAO} a short note with a non-stochastic labor income is considered.

\subsection{The annuity market}

Consider an individual aged $\chi$ at time $0$. We shall denote by $\psg{s-t}{\chi+t}$ the subjective conditional probability that an individual aged $\chi+t$ believes she will survive at least $s-t$ years (i.e. to age $\chi+s$). We  recall that $\psg{s-t}{\chi+t}$ can be defined through the force of mortality. Let $F_{\chi+t}(s)$ denotes the cumulative distribution function of the time of death of an individual aged $\chi+t$. Assuming that $F_{\chi+t}$ has the probability density $f_{\chi+t}$, the force of mortality at age $\chi+t+\eta$ is defied by
        \[\lambda^S_{\chi+t}(\eta) \deq \frac{f_{\chi+t}(\eta)}{1-F_{\chi+t}(\eta)},\]
which leads to
        \[\psg{s-t}{\chi+t} = e^{-\int_0^{s-t} \lambda^S_{\chi+t}(\eta) \de \eta}\]

Since  it is possible to obtain $\lambda^S_{\chi+t}(\eta) = \lambda^S_{\chi+t+\eta}(0)$ (see Gerber \cite{Gerber_LIB}), it is useful to denote $\lambda^S_{\chi+t}(\eta)$ by the symbol $\lambda^S_{\chi+t+\eta}$. This leads to write
        \beq \psg{s-t}{\chi+t} = e^{-\int_t^s \lambda^S_{\chi+\eta} \de \eta}
        \label{eq:tPx_and_lambda}\eeq

For the numerical simulations below, we will assume a Gompertz's specification for the force of mortality $\lambda^S$:
\[\lambda^S_{\chi+\eta} \deq  \frac{1}{\varsigma} \exp\tonde{\frac{\chi+\eta - m}{\varsigma}}\]

Similar formulas are given for both the objective conditional probability of survival $\pog{s-t}{\chi+t}$ and the objective hazard function $\lambda^O$. Employing the method proposed by Carriere \cite{Carriere94_UltimateParametricModel}, we estimate the parameters $m$ and $\varsigma$ in Table \ref{tb:GompertzEstimations} using the Human Mortality Database for the province of Ontario, Canada, for a female and a male both aged 35 in the years 1970 and 2004.

\begin{table}[h]
    \caption{Estimated Gompertz's parameters for a female and a male from the province of Ontario, Canada, conditional on survival to age 35.  Source: Canadian Human Mortality Database available for year 1970 and 2004. }\label{tb:GompertzEstimations}
    \begin{center}
    \begin{tabular}{l c r r c r r}
                    \cline{1-7}
                   &    & \multicolumn{2}{c}{\footnotesize Female} &  & \multicolumn{2}{c}{\footnotesize Male} \\
                   \cline{3-7}
                    \multicolumn{1}{c}{\footnotesize Year} &    &  \multicolumn{1}{c}{\footnotesize $m$}  & \multicolumn{1}{c}{\footnotesize $\varsigma$}   &   &     \multicolumn{1}{c}{\footnotesize $m$}  & \multicolumn{1}{c}{\footnotesize $\varsigma$}\\
                    \cline{1-1} \cline{3-4} \cline{6-7}\addlinespace[2pt]
                    \footnotesize1970 &  & \footnotesize 85.3758 & \footnotesize 10.5098  &  & \footnotesize 79.1089 & \footnotesize 11.5890\\
                    \footnotesize2004 &  & \footnotesize 89.7615 & \footnotesize 9.3216    &  & \footnotesize 85.8651 & \footnotesize 10.1379\\
                    \addlinespace[2pt]\cline{1-7}
    \end{tabular}
    \end{center}
\end{table}

In a continuous compounding setting with a constant interest rate $r$,  the (present) actuarial value of a life annuity that pays  at unit rate per year for an individual who is age $\chi+t$ at time $t$, is given by
        \[\ol a_{\chi+t} \deq \int_t^{+\infty} e^{-r(s-t)}\pog{s-t}{\chi+t} ds\]
where the survival probability is determined considering the objective mortality assessment from the insurer's point of view.

For a given $\ol a_{\chi+t}$,  an individual endowed by a wealth $x>0$ is able to buy $x/\,\ol a_{\chi+t}$ unit rate annuities, corresponding to a cash--flow stream at a nominal  instantaneous rate $H\deq x/\,\ol a_{\chi+t}$. This defines a conversion rate $h\deq 1/\,\ol a_{\chi+t}$, at which an amount $x$ can be turned into a life long annuity with an income stream of $H = x h$ per annum.

Notice that the conversion rate $h$ imply a technical nominal instantaneous rate $r_h$ defined by the following expression:
        \[\frac{1}{h} = \ol a_{\chi+t}^{\,(h)}\deq  \int_t^{+\infty} e^{-r_h(s-t)}\pog{s-t}{\chi+t} ds\]
which depends on the mortality assumptions summarized by $p^O$.

\begin{figure}[h]
  \caption{Simulated implicit rate $r_h$, assured by the an insurer in 1970 with respect to different values for the guaranteed conversion rate $h$. The policyholder is supposed to be a 35 years old female, from the province of Ontario, who will be 65 at the time $T$ of retirement. Values for $r_h$ are compared (with an approximation of 1E-04) using a Gompertz's mortality function with (objective) parameters driven by survival tables available in 1970 (solid line) and in 2004 (dashed line). }
  \begin{center}
  \includegraphics[scale=0.32]{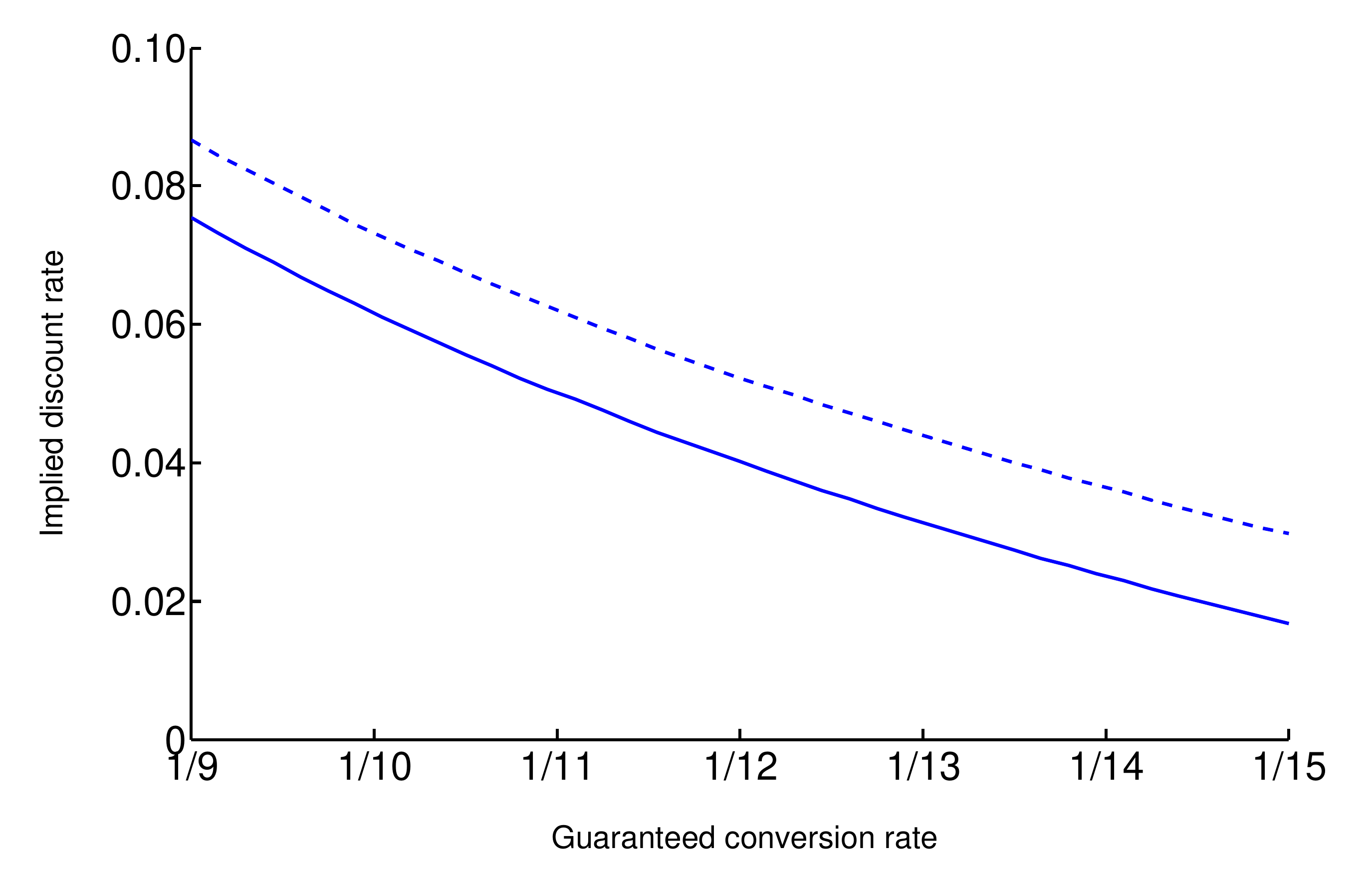}\\   %scale = 0.32 (PDF) 0.40(PDF->EPS)
  \end{center}
  \label{fig:rh_wrt_h}
\end{figure}

Returning to our \gao policy, in order to offer a given conversion rate $h$ to be used by the policy holder at time $T$, the insurer considers the interest and mortality rates based on information available at time $t_0$. However, improvements in mortality rates and the decline in market interest rates may represent an important source of liabilities for the insurer. For instance, if at time $T$, the interest rates will be below the technical rate $r_h$ and the policyholder decides to exercise the guaranteed annuity option, the insurer has to make up the difference between the two rates. Figure \ref{fig:rh_wrt_h} plots the implicit rate $r_h$ with respect to different  values for the conversion rate of $h$. The same figure also shows the impact of the so called \textit{longevity risk}: taking $h=1/9$ (very common in 1970's and 1980's) a rate of $r_h = 0.0754$ is implicitly determined, assuming a mortality specification which was available from estimations in 1970.  However $r_h$ rises to $0.0867$, when the estimation of $p^O$ is made by the mortality tables available in 2004. Hence, as remarked by Boyle \& Hardy \cite{BoyleHardy_WP_GuarnAnnOpts}, if mortality rates improve so that policyholders live systematically longer, the interest rate at which the guarantee becomes effective will increase.

%=====================================================
    \subsection{The valuation method}
    \label{subsec:MainConsiderations}
%=====================================================

As outlined in the introduction, our valuation method will be based on two steps, motivated by the following two remarks:
\begin{enumerate}
        \item provided the guaranteed annuity option has been purchased at time $t_0=0$, the policyholder needs to decide on whether or not to exercise it at time $T$;
        \item assuming an optimal exercise decision at time $T$, the policyholder needs to decide how much she would pay at time $t_0$ to embed such an option in her policy.
\end{enumerate}

In order to obtain a well-posed valuation for this option, we need to assume that the purchased insurance policy does not include any other guarantees or rights. We also assume that the conversion period (i.e. the time interval in which the agent is asked to take a decision on whether to exercise the option) reduces to the instant $T$.

\subsection{Optimal exercise at time $T$}
\label{subsec:AnalysisOption_T}
%--------------------------------------------------------------
\vspace{0.1in}

At time $T$, if the policyholder owns a guaranteed annuity option, she will be asked to take the decision to convert the accumulated funds in a life long annuity at the guaranteed rate, or to withdraw the money and invest in the market. Therefore, we need to compare the following two strategies:

\begin{itemize}

    \item[\textit{i})] If she decides to convert her accumulated funds $A>0$ at the pre-specified conversion rate $h$, she will receive a cash flow stream at a rate $H = A\cdot h$ per annum. In this case, we assume that, from time $T$, she will be allowed to trade in the financial market. Henceforth,  her instantaneous income will be given by the rate $H$ and by the gains she is able to realize by trading in the financial market.

    \item[\textit{ii})] On the other hand, if the policyholder decides not to convert her funds into the guaranteed annuity, we assume she can just withdraw funds $A$ and  go in  the financial market. In this case, from time $T$, her income will be represented just by the market gains she can realize. Her total endowment at the future time $T$ will be then increased by the amount $A$.
\end{itemize}

Since the accumulation phase regards the period $[t_0,\,T)$, we assume that just before the time $T$ the policyholder's wealth is given by $X_{T_-}=x_T>0$. If she decides to convert her accumulated funds exercising the guaranteed annuity option, the problem she will seek to solve is to maximize the present value of the expected reward represented by value function $V$ defined as follows:
        \[V(x_T,\,T)\deq \sup_{\graff{c_s,\,\pi_s}}\E\left[\int_T^{+\infty} e^{-r(s-T)} \psg{s-T}{\chi+T} \right. \cdot  \left. \left.  \vphantom{\int_T^{+\infty}} u(c_s) \de s \ \right| X_T = x_T \right]\]
where the function $u$ is the policyholder's utility of consumption, which is assumed to be twice differentiable, strictly increasing and concave, and the wealth process $X_t$ satisfies, for all $t\geq T$, the dynamics
\begin{eqnarray}
dX_t &=& r (X_t -\pi_t)dt +\pi_t(dS_t/S_t) + (H-c_t)dt, \nonumber \\
        &=& \quadr{rX_t + (\mu - r)\pi_t +H-c_t }dt + \sigma \pi_t dW_t,
        \label{wealth_T}
\end{eqnarray}
with initial condition $X_T=x_T$. On the contrary, if the policyholder decides not to exercise the option, she withdraws the accumulated funds $A$ at time $T$ and to solve a standard Merton's problem given by:
       \[U(x_T+A,\,T)\deq \sup_{\graff{c_s,\,\pi_s}}\E\left[\int_T^{+\infty} e^{-r(s-T)} \psg{s-T}{\chi+T}
        \left. \vphantom{\int_T^{+\infty} } u(c_s) \de s \ \right| X_T = x_T + A \right]\]
with initial condition $X_T=x_T+A$ and subject to the following dynamics:
\begin{eqnarray}
dX_t &=& r (X_t -\pi_t)dt +\pi_t(dS_t/S_t) -c_t dt, \nonumber \\
        &=& \quadr{rX_t + (\mu - r)\pi_t -c_t }dt + \sigma \pi_t dW_t,
        \label{wealth_T_noH}
\end{eqnarray}

We assume the control processes $c_t$ and $\pi_t$ are admissible, in the sense that they are both progressively measurable with respect to the filtration $\{\F_t\}$. Also, the following conditions hold a.s. for every $t\geq T$:
      \begin{equation}
      \label{admiss}
      c_t\geq 0, \quad \int_{T}^t c_s ds < \infty \quad \text{and } \int_{T}^t \pi_s^2 ds <\infty
      \end{equation}

At time $T$ the policyholder compares the two strategies described above and the respective expected rewards. We postulate she will decide to exercise the guaranteed annuity as long as
        \[U(x_T + A,\,T )\leq V(x_T,\,T)\]

The previous analysis, regarding the function $U$, considers a policyholder that holds a policy embedding a guaranteed annuity option. A third strategy needs to be considered in order to describe the case in which the policyholder holds a policy with no guaranteed annuity option embedded in it:

\begin{itemize}
    \item[\textit{iii)}] If the policy does not embed a guaranteed annuity option, the policyholder does not have the right to convert the accumulated funds $A$ into a lifelong annuity. In this sense, we assume that value function $U$ will represent the expected reward if at time $t_0$ the policyholder purchased a plan without the guaranteed annuity option.
\end{itemize}

\subsection{Optimal strategies at time $t_0$}
\label{subsec:AnalysisOption_t0}
%-----------------------------------------------------------------
After defining the optimal exercise at time $T$, we can formalize the policyholder's analysis at the initial time $t_0$, when the guaranteed annuity option may be embedded in her policy. We can summarize the two strategies that the policyholder faces at time $t_0$ as follows:

\begin{itemize}
    \item[\textit{i})] the policyholder purchases a policy without embedding a guaranteed annuity option in it. In this case, she will pay a continuous premium at an annual rate $P$ to accumulate funds $A$ up to time $T$;
    \item[\textit{ii})] the policyholder decides to embed the guaranteed annuity option in her policy. She will pay a lump sum $L_0$ for this extra benefit immediately and the continuous premium $P$ for the period $[t_0,\,T)$, as in previous case.
\end{itemize}

In either case, the value of the accumulated funds is given by
        \[A=\int_{t_0}^T e^{r(T-s)} Pds\]

Assuming the policyholder's income is given by the  gains she can realize trading in the stock market, between time $t_0$ and time $T$, her wealth needs to obey to the following dynamics:
        \beq\begin{split}
        dX_t &= r(X_t -\pi_t)dt + \pi_t (dS_t/S_t) - (P+c_t)dt\\
        &= [rX_t + (\mu - r)\pi_t - P - c_t]dt + \sigma \pi_t dW_t\end{split}\label{eq:Dynamics_calUV}\eeq
with initial condition $X_{t_0}= x_0>0$. Therefore, if the agent decides not to embed the guaranteed annuity option in her policy she will seek to solve the following optimization problem
        \begin{multline*}\cal U (x_0,\,t_0) \deq \sup_{\{c_s,\,\pi_s\}} \E \left[\int_{t_0}^T e^{-r(s-t_0)}\psg{s-t_0}{\chi+t_0}\cdot u(c_s)  ds \right. +\\
        +\left.\left. \vphantom{\int_{t_0}^T} e^{-r(T-t_0)}\psg{T-t_0}{\chi+t_0}\cdot U(X_T + A,\,T) \ \right| X_{t_0}=w_0 \right]\end{multline*}

On the contrary, if she decides to embed a \gao in her policy, paying the lump sum $L_0$ at time $t_0$,  her wealth is still given by dynamics (\ref{eq:Dynamics_calUV}), but the maximization problem will be different, namely for $w_0-L_0>0$:
        \begin{multline*}\cal V (x_0-L_0,\,t_0) \deq \sup_{\{c_s,\,\pi_s\}} \E \left[\int_{t_0}^T e^{-r(s-t_0)}\psg{s-t_0}{\chi+t_0}\cdot u(c_s) \de s \right. +\\
        +\vphantom{\int^T_{t_0}} e^{-r(T-t_0)}\psg{T-t_0}{\chi+t_0} \left.\left. \vphantom{\int_{t_0}^T} \negthickspace\! \max\bgraff{U(X_T + A,\,T;r),\ V(X_T,\,T)} \ \right| X_{t_0}=x_0 - L_0 \right]\end{multline*}
Notice that at time $T$ the policyholder can an either exercise the option, remaining with the wealth $X_T$ plus the lifelong annuity obtained from converting $A$ at the rate $h$, or decide not to exercise the option and withdraw the accumulated funds $A$.

The same admissibility conditions are required for the control processes $c_t$ and $\pi_t$ during the accumulation period, namely
they are both progressively measurable with respect to $\F_t$ and satisfy \eqref{admiss}

We postulate that the agent will decide to embed a guaranteed annuity option in her policy as long as the following inequality holds:
        \[\cal U(x_0,\,t_0)\leq \cal V(x_0 - L_0, t_0 )\]

\subsection{The inequality $U(X_T+A,\,T) \leq V(X_T,\,T)$}
%-------------------------------------------------------------
Given a wealth $X_T$ at time $T$, Grasselli \& Silla \cite[App. A]{GrasselliSilla_BuyerGAO} find an explicit solution for a class of problems regarding value functions $U$, $V$, $\cal U$ and $\cal V$, assuming a constant relative risk aversion (CRRA) utility function defined by
        \beq u(c)=\frac{c^{1-\gamma}}{1-\gamma}, \quad \gamma >0,\ \gamma\neq 1
        \label{eq:UtilityFunction_u}\eeq
and a constant interest rate satisfying the condition $r>(1-\gamma)\delta$ where
\[\delta\deq r+1/(2\gamma)\cdot (\mu - r)^2/\sigma^2.\]
Namely,   the value functions $U$ and $V$ are given by:
        \begin{align}
        U(X_T+ A,\,T) &= \frac{1}{1-\gamma}\tonde{X_T + A}^{1-\gamma} \cdot \phi^\gamma(T) \\
        V(X_T,\,T) &= \frac{1}{1-\gamma}\tonde{X_T + \frac{H}{r}}^{1-\gamma} \cdot \phi^\gamma(T)
        \end{align}
where $\phi$ is is given by
    \beq \phi(T) = \int_T^{+\infty} e^{-b(s-T)}\cdot \psg{s-T}{\chi+T}ds
    \label{eq:function_phi}\eeq
for $b\deq -\quadr{(1-\gamma)\delta - r}/\gamma$. Notice that for every $\gamma >0$, $\gamma \neq 1$, we have
        \[U(X_T+A,\,T) \leq \ V(w_T,\,T)\quad \iff \quad r\leq h\]

From an economic point of view,  the previous inequality tells us that, at time of conversion $T$, the policyholder will find convenient to exercise the guaranteed annuity option if and only if the guaranteed rate $h$ is greater than the prevailing interest rate $r$. Moreover, recalling that $1/h = \ol a_{\chi+T}^{\,(h)}$, the previous inequality can be also written as follows:
        \[U(X_T+A,\,T) \leq \ V(X_T,\,T)\quad \iff \quad \ol a_{\chi+T}^{\,(h)}\leq 1/r,\]
which says that in order to come to a decision the policyholder compares the guaranteed cost of a unit rate lifelong annuity (assured by the insurance company), whose the present value is given by a guaranteed implicit rate $r_h$, with the market cost of a unit rate perpetuity, whose present value is determined by the market interest rate $r$. Notice that the indifference point is given by $\ol a_{\chi+T}^{\,(h)} = 1/r$, highlighting the absence of bequest motives for the policyholder after time $T$.

\subsection{A closed form for value functions $\cal U$ and $\cal V$}
%--------------------------------------------------------------

Combining the results of the previous two sections, we have that the value function $\cal U$ at time $T$ needs to be equal to
\[g(X_T)=\frac{(X_T + A)^{1-\gamma}}{1-\gamma}\phi^\gamma(T).\]
Using the change of variables technique proposed in Grasselli \& Silla \cite[App. B]{GrasselliSilla_BuyerGAO}, we find that the value function $\cal U$ is given by
        \begin{equation}
        \cal U(x_0,\,t_0) = \frac1 {1-\gamma} \tonde{x_0 - \wh \xi_{\cal U} (t_0) }^{1-\gamma}\phi^\gamma(t_0)
        \end{equation}
where $\wh \xi_{\cal U}$ is defined by
    \beq \wh \xi_{\cal U}(t) = \frac{P}{r} \tonde{1-e^{r(t-T)}} - A \cdot e^{r(t-T)}.
    \label{eq:xi_hat_U}\eeq

Similarly, we have that the value function ${\cal V}$, at time $T$, needs to be equal to
        \begin{eqnarray*}
        G(x_T)&\deq& \max\bgraff{U(x_T+A,\,T), \ V(x_T,\,T)}\\
        &=& \left\{\!\!\begin{array}{ll}
                               \frac{(X_T + A)^{1-\gamma}}{1-\gamma}\phi^\gamma(T), & \qquad \text{if}  \ \ r \geq h \vspace{1ex}\\
                               \frac{(X_T + H/r)^{1-\gamma}}{1-\gamma}\phi^\gamma(T),       & \qquad \text{if} \ \  r < h
                              \end{array}
        \right.\end{eqnarray*}
Using the same change of variables technique, we arrive at the following expression for the value function $\cal V$:
        \[\cal V(w_0,\,t_0) = \left\{\!\!\begin{array}{ll}
                                       \cal U(w_0,\,t_0) & \text{  if} \ \ r \geq h  \vspace{1ex}\\
                                        \frac1 {1-\gamma} \tonde{w_0 - \wh \xi_{\cal V} (t_0) }^{1-\gamma}\phi^\gamma(t_0) & \text{  if} \ \ r < h
                                     \end{array}
        \right.\]
where $\wh \xi_{\cal V}$ is given by
    \beq \wh \xi_{\cal V} (t) = \frac P r \tonde{1-e^{r(t-T)}}-\frac H r \cdot e^{r(t-T)}
    \label{eq:xi_hat_V}\eeq

\subsection{The indifference valuation for the guaranteed annuity option}
%-------------------------------------------------------------
Consider the policyholder that, at time $t_0$, compares the two expected rewards  arising from the value functions $\cal U$ and $\cal V$, and define the  indifference value for the guaranteed annuity option by
        \[ L^*_0\deq\sup\bgraff{ L_0 : \cal U(w_0,\,t_0) \leq \cal V(w_0 -L_0,\,t_0), \ w_0 - L_0 >0}\]
If the indifference value exists, it is straightforward to deduce that it is given by
        \[L^*_0 = \tonde{\frac{H}{r} - A} e^{-r(T-t_0)}\]

\subsection{Stochastic interest and mortality rates}
        \label{sec:Extension_StochasticHazardRate}

As we mentioned in the introduction, the liabilities associated with guaranteed annuity options depends on the variations of interests rates and mortality rates over the time. In this sense a richer model has to take into account and to formalize these rates as stochastic processes. In present section we will just offer a sketch for stochastic models for mortality intensity.

The debate over the stochastic mortality is very prolific and the literature concerning this problem is huge. For what concerns in particular mortality trends and estimation procedure, we recall for example: Carriere \cite{CARRIERE94_InvestigationOfGompertzLaw} and \cite{Carriere94_UltimateParametricModel}, Frees, Carriere and Valdez \cite{FreesCarriereValdez_ART_AnnuityValuation}, Stallard \cite{Stallard__Demographic}, Willets \cite{Willets_ART_MortalityEmprovement} and \cite{Willets_ART_LongevityIn21Century}, Macdonald et al. \cite{Macdonald_ART_InternationalComparisonMortalityTrends} and Ruttermann \cite{Ruttermann_ART_MortalityTrendsWorldwide}. In what concern stochastic diffusion processes to model the force of mortality, excellent contributions are offered by: Lee \cite{Lee_ART_TheLeeCarterMethod}, Pitacco \cite{Pitacco_ART_SurvivalModels}, \cite{Pitacco___ART_LongevityRiskInLivingBenefits}, Olivieri and Pitacco \cite{OlivieriPitacco_ART_AnnuitisationGuaranteeAndUncertainty}, \cite{OlivieriPitacco_ART_ForecastingMortality}, \cite{OlivieriPitacco_ART_RenditeVitalizeLongevityRisk}, Olivieri \cite{Olivieri_ART_UncertaintyInMortalityProgections}, Dahl \cite{Dahl_ART_StochasticMortality}, Schrager \cite{Schrager_ART_AffineStochasticMortality}, Cairns et al. \cite{CairnsBlakeDowd_ART_ATwoFactorModel}, Marceau Gaillardetz \cite{MarceauGaillardetz}, Milevsky, Promislow and Young \cite{Milevsky_Promislow_Young_2005}, \cite{Milevsky_Promislow_Young_2006}. We also recall that the approach followed by Milevsky and Promislow \cite{Milevsky_Promislow_2001} was the pioneering contribution that consider at one time both the stochastic mortality and a financial market model, in order to price the embedded option to annuitise (what we call guarantee annuity option).

The contribution by Dahl \cite{Dahl_ART_StochasticMortality}, propose to model the mortality intensity by a fairly general diffusion process, which include the mean reverting model proposed by Milevsky and Promislow \cite{Milevsky_Promislow_2001}. Precisely the author consider a $\P$ dynamics for the mortality intensity given by
        \beq \de \lambda_{\chi+s} = \alpha^\lambda \tonde{s,\,\lambda_{\chi+s}}\de s + \sigma^\lambda\tonde{s,\,\lambda_{\chi+s}}\de \wt W_s\label{eq:Dynamics_for_lambda}\eeq
where $\alpha^\lambda$ and $\sigma^\lambda$ are non-negative and $\{\wt W_s\}$ is a standar Wiener process with respect to the same filtration $\graff{\F_s}$, defined above, for $s\geq t_0$. $\{\wt W_s\}$ is assumed uncorrelated with $\{W_s\}$.

In order to avoid analytical difficulties, we investigate the effect of varying mortality rates by comparing different scenarios for different survival probabilities. In particular, the next section highlights the effect of different parameterized functions describing different specifications concerning the force of mortality.

\section{Numerical examples and insights}
\label{sec:Numerics_Insights}

\subsection{Valuation under different scenarios interest rate scenarios}
\label{subsec:NumericsOver_r}

Consider $t_0 = 0$ and, at this time, a female aged $\chi = 35$ who is willing  to purchase a policy. Also, suppose that this plan will accumulate, until time $T \deq 30$ (i.e. when the policyholder will be aged $\chi+T=65$) an amount $A \deq \text{\$}350,000$. In order to be concrete, we can think that $T$ may coincide with her retirement time and  that the purchase takes place in 1975. In this context,  the \gao (if the agent decides to embed such an option in her policy) could be exercised in 2005. We would like to stress that these calendar dates are not necessary to implement a numerical experiment. However they give a stronger economic meaning for  a contract designed as follows: we assume that  the agent is asked to decide whether to include  a guaranteed annuity option with a conversion rate $h\deq 1/9$ (very common in 1980's and 1970's), implying a guaranteed cashflow stream at the nominal rate $H \approx \$38,888.89$ per year. Notice that, in this situation, if we refer to survival tables available in 1970 (see table \ref{tb:GompertzEstimations}), the implicit discount rate is $r_h \approx 0.0754$ and such an option was considered to be far out-of-the-money for the policyholder.

\begin{figure}[t]
  \caption{Value function $\cal U$ (solid) and value function $\cal V$ (dashed), for an individual characterized by $\gamma = 1.4$, that observes a financial market described by $r = 0.07$, $\mu = 0.08$, $\sigma=0.12$. The value of $r$ and $\mu$ are taken large enough to simulate the 1970's financial market. In this setting we find $L^*_0 = 25,171$. The price is given for a \gao exercisable in 2005, for a female in year 1970, from the province of Ontario,  assuming a (subjective) mortality specification given by the survival table available in 1970, see Table \ref{tb:GompertzEstimations}.}
  \begin{center}
  \includegraphics[scale=0.32]{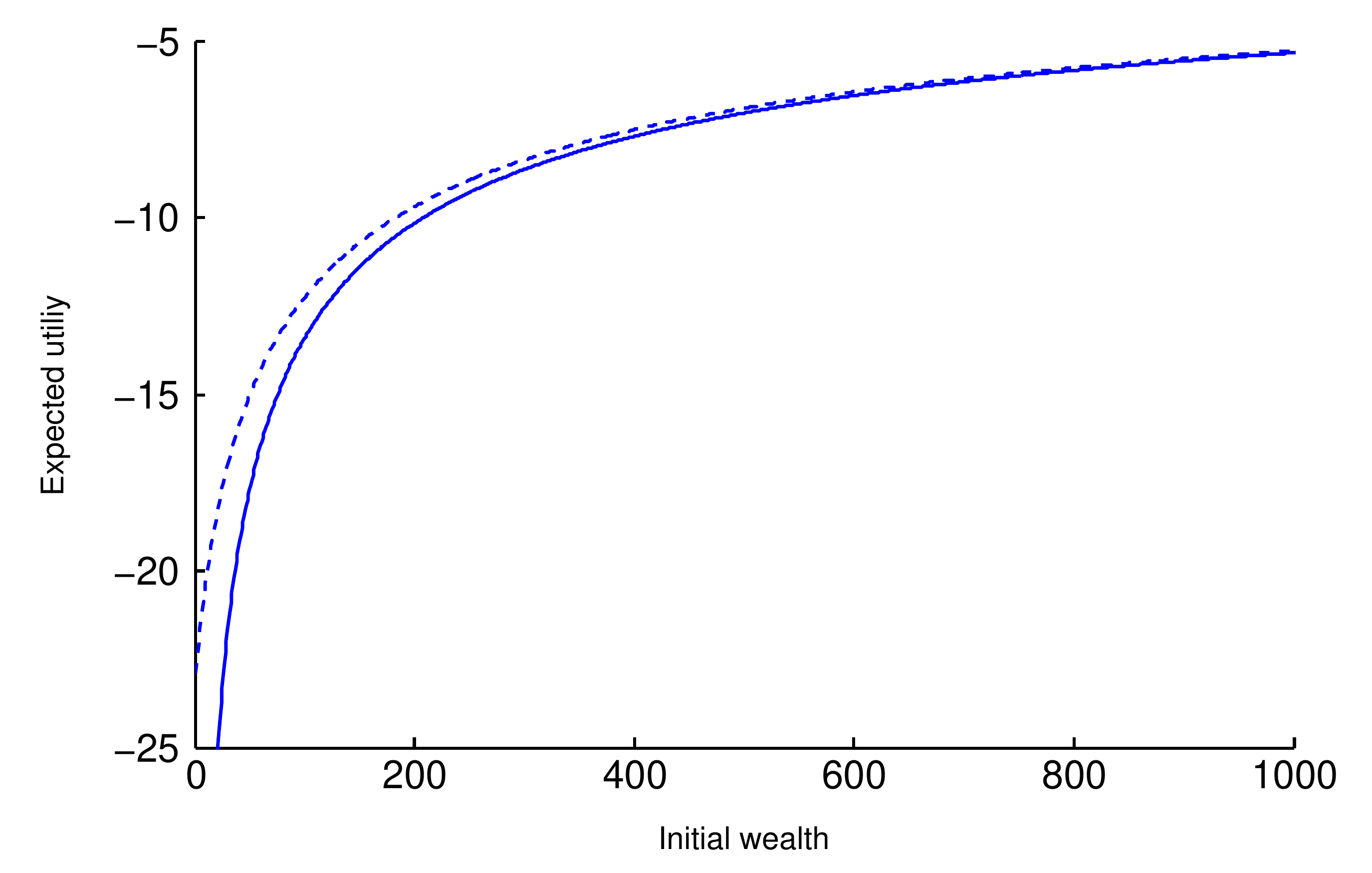}
  \end{center}
  \label{fig:ValFunct_calUV_a}
\end{figure}

Under the previous hypothesis, the value functions $\cal U$ and $\cal V$ are plotted in figure \ref{fig:ValFunct_calUV_a}, where  we assume a Gompertz's mortality specification. We estimate the parameters $\varsigma$ and $m$, minimizing a loss function using the method proposed by Carriere \cite{Carriere94_UltimateParametricModel}. We refer to the Human Mortality Database for the province of Ontario, Canada,  for a female and a male both aged $35$ using tables available in 1970 or in 2004. The results of our estimations are  summarized in  Table \ref{tb:GompertzEstimations}.

\begin{figure}[ht!]
  \caption{Indifference price $L^*_0$ depending on the guaranteed conversion rate $h$ and the market interest rate $r$. The valuation is given for a \gao exercisable in 2005, for a female in year 1970, from the province of Ontario,  assuming a (subjective) mortality specification given by the survival table available in 1970, see Table \ref{tb:GompertzEstimations}.}\vspace{0ex}
  \begin{center}
  \includegraphics[scale=0.37]{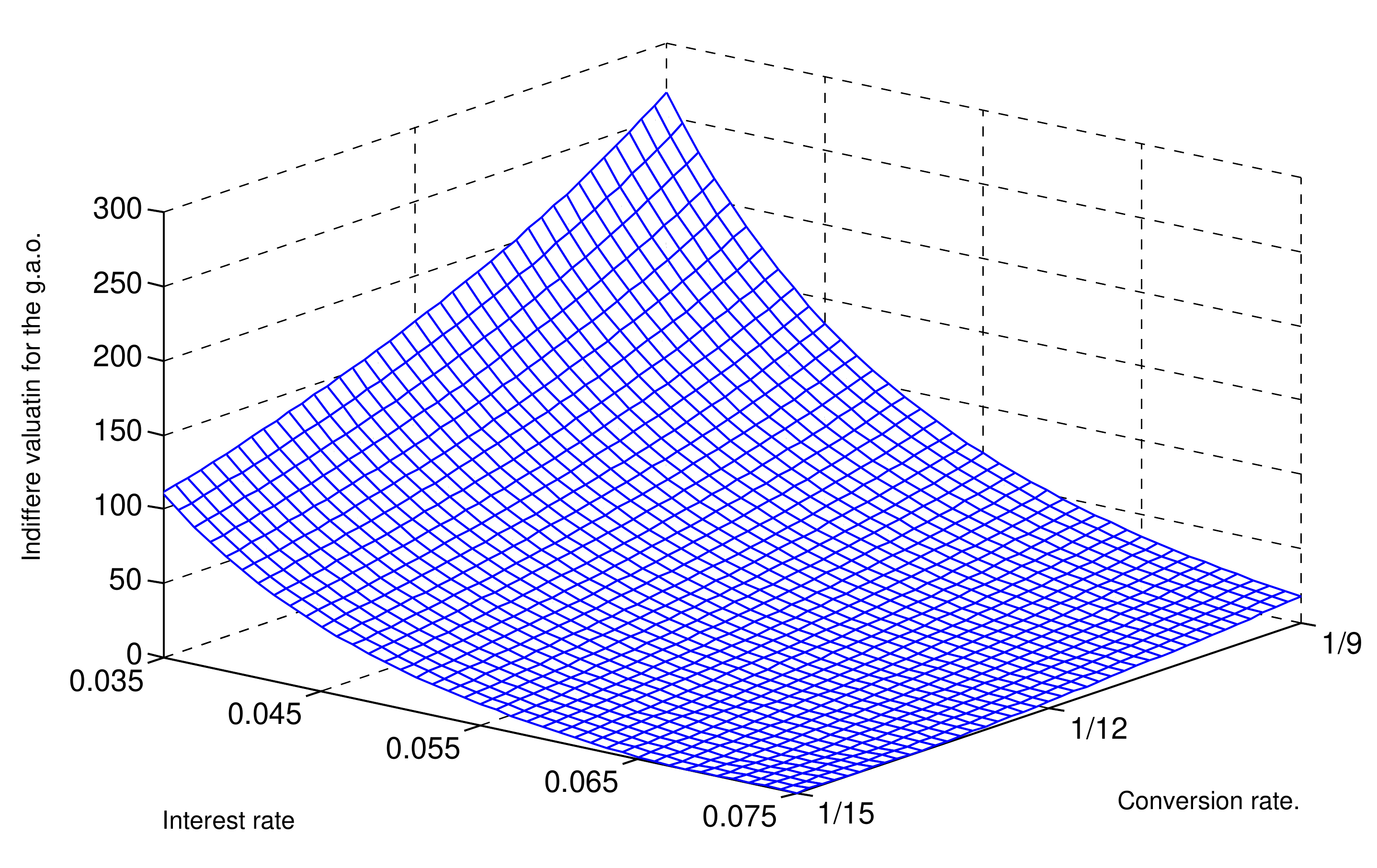}
  \end{center}
  \label{fig:ValFunct_calUV_Lhr}
\end{figure}

For some values of the market interest rate $r$, Table \ref{tb:MontlyPremiums} shows the premium $P$ and the equivalent valuation $L^*_0$ for this policy. Figure \ref{fig:ValFunct_calUV_Lhr} depicts the dependency of $L^*_0$ on both the guaranteed conversion rate $h$ and the interest rate $r$. As expected, the greater the interest rate, the lower the policyholder's indifference price for the option. Also, the analysis remains consistent with respect to  $h$: the lower the guaranteed rate, the lower the agent's indifference price.

\begin{table}[t]
    \caption{Premium and indifference valuation associated to the policy, depending on the current interest rate.}
    \label{tb:MontlyPremiums}
    \begin{center}
    \begin{tabular}{l c r r c r r r}
                    \cline{1-8}
                    \multicolumn{1}{c}{\footnotesize $r$}&    & \multicolumn{1}{c}{\footnotesize $P$} &\multicolumn{1}{c}{\footnotesize $L^*_0$}  &  & \multicolumn{1}{c}{\footnotesize $p_{12}$} & \multicolumn{1}{c}{\footnotesize $l_{12}$} & \multicolumn{1}{c}{\footnotesize Total}\\
                   \cline{1-1}\cline{3-4}\cline{6-8}\addlinespace[2pt]
                    \footnotesize 0.035 &  & \footnotesize \$6,594 & \footnotesize \$266,342                             &  & \footnotesize \$550 & \footnotesize \$419 & \footnotesize \$969\\
                    \footnotesize 0.050 &  & \footnotesize \$5,026 & \footnotesize \$\phantom{0}95,450    &  & \footnotesize \$420 & \footnotesize \$115  & \footnotesize \$535\\
                    \footnotesize 0.085 &  & \footnotesize \$2,519 & \footnotesize \$\phantom{00}8,395    &  & \footnotesize \$211 & \footnotesize \$\phantom{00}5  & \footnotesize \$216\\
                    \addlinespace[2pt]\cline{1-8}
    \end{tabular}
    \end{center}
\end{table}

Depending on $r$, Table \ref{tb:MontlyPremiums}, shows the nominal instantaneous rate for the premium $P$ (that the policyholder needs to pay to in order to accumulate $A = \$350,000$) and the indifference valuation $L^*_0$ for the \gao.  Notice that it is not immediately possible to compare $L^*_0$ and $P$ since the former denotes a lump sum, while the latter refers to a nominal instantaneous rate to be paid over time.

In order to better understand the meaning of $P$ and $L^*_0$, it can be useful to think of an auxiliary problem. This problem is independent of the previous indifference model, but will offer a way to validate the previous results. To do this, consider a premium to be payed monthly for a pension or an insurance plan. We can ask two questions: what is the value $p_{12}$ of a monthly payment whose the future value, after 30 years, is exactly $A$; and what is the value of a monthly payment $l_{12}$ necessary to amortize, after 30 years,  the lump-sum $L^*_0$ payed at $t_0=0$.

In order to compute $l_{12}$, consider a horizon of $T\per 12$ months. Thus $l_{12}$ is given by the following relation:
        \[L^*_0 = l_{12}\cdot \afig{T\per 12}{i_{12}}\]
where $i_{12} \deq  e^{r/12} - 1$ is the effective interest rate compounded monthly with respect to $e^r$, and where in general we define
        \[\afig{n}{i} \deq \frac{1- (1+i)^{-n}}{i}\]
 as the present value of an annuity that pays one dollar for $n$ periods, discounted by the effective interest rate $i$ compounded each period. Similarly, define $p_{12}$ such that
        \[A = p_{12}\cdot \sfig{T\per 12}{i_{12}}\]
where
        \[\sfig{n}{i} \deq \frac{(1+i)^{n}-1}{i} = (1+i)^n\cdot \afig{n}{i}\]
represents the future value after $n$ periods, of an annuity that pays one dollar per period, under an effective interest rate $i$ compounded each period.

Coming back to Table \ref{tb:MontlyPremiums} it is interesting to see that for $r = 0.035$, a monthly cash flow of \$550 and a monthly stream of \$419
equivalently amortize $L^*_0$.  Setting $r=0.085$, we observe a similar situation for a monthly premium of \$211 and a monthly stream of only \$5.
These intuitive results are consistent with the literature concerning the guaranteed annuity option. As mentioned by Boyle \& Hardy \cite{BoyleHardy_ART_GuaranAnnOpts},  these guarantees were popular in U.K. retirement savings contracts issued in the 1970's and 1980's, when long-term interest rates were high. The same authors also write that at that time, the options were very far out--of--the-- money and insurance companies apparently assumed that interest rates would remain high and thus the guarantees would never become active. As a result, from the indifference model discussed in the present paper, when the interest rate is very high - as was the case in the 1970's and 1980's - the guaranteed annuity option's value, from the point of view of the policyholder, is very small. Interestingly, in the same period, empirically it was observed that a very small valuation was also given by insurers.

These facts are confirmed by the extremely low value of $L^*_0 = \$8,395$ (over $T-t_0= 30$ years), against the yearly nominal premium  $P = \$2,519$. This is better seen in terms of the auxiliary \textv{monthly valuation problem}: the lump sum $L^*_0$ can be amortized by a monthly cash flow of \$5, against a monthly equivalent premium of \$211.  Moreover,  $p_{12}$ and $l_{12}$ by construction are homogeneous quantities.  Their sum  gives an idea of the equivalent monthly value associated to the policy the agent is willing to buy at time $t_0$. This sum is showed in the last column of Table \ref{tb:MontlyPremiums}. It is interesting to note the large difference between the total value corresponding to $r = 0.035$ compared to $r = 0.085$.

\subsection{Valuation under different mortality scenarios}

\begin{figure}[h]
  \caption{Value function $\cal U$ (solid) and value function $\cal V$ (dashed), for a guaranteed annuity option maturing in 2005, for a 35 years old female in year 1975, from the province of Ontario,  comparing a (subjective) mortality specification given by the survival table available in 1970 (light lines) and in 2004 (demi-bold lines), see Table \ref{tb:GompertzEstimations}. The policyholder is characterized by $\gamma = 1.4$, observing a financial market described by $r = 0.07$, $\mu = 0.08$, $\sigma=0.12$. The value of $r$ and $\mu$ are taken large enough to simulate the 1970's financial market.}
    \label{fig:StochMortality}
  \begin{center}
 \subfigure[Value functions $\cal U$ and $\cal V$.]{\label{fig:StochMortality_ValFunc}\includegraphics[scale=0.33]{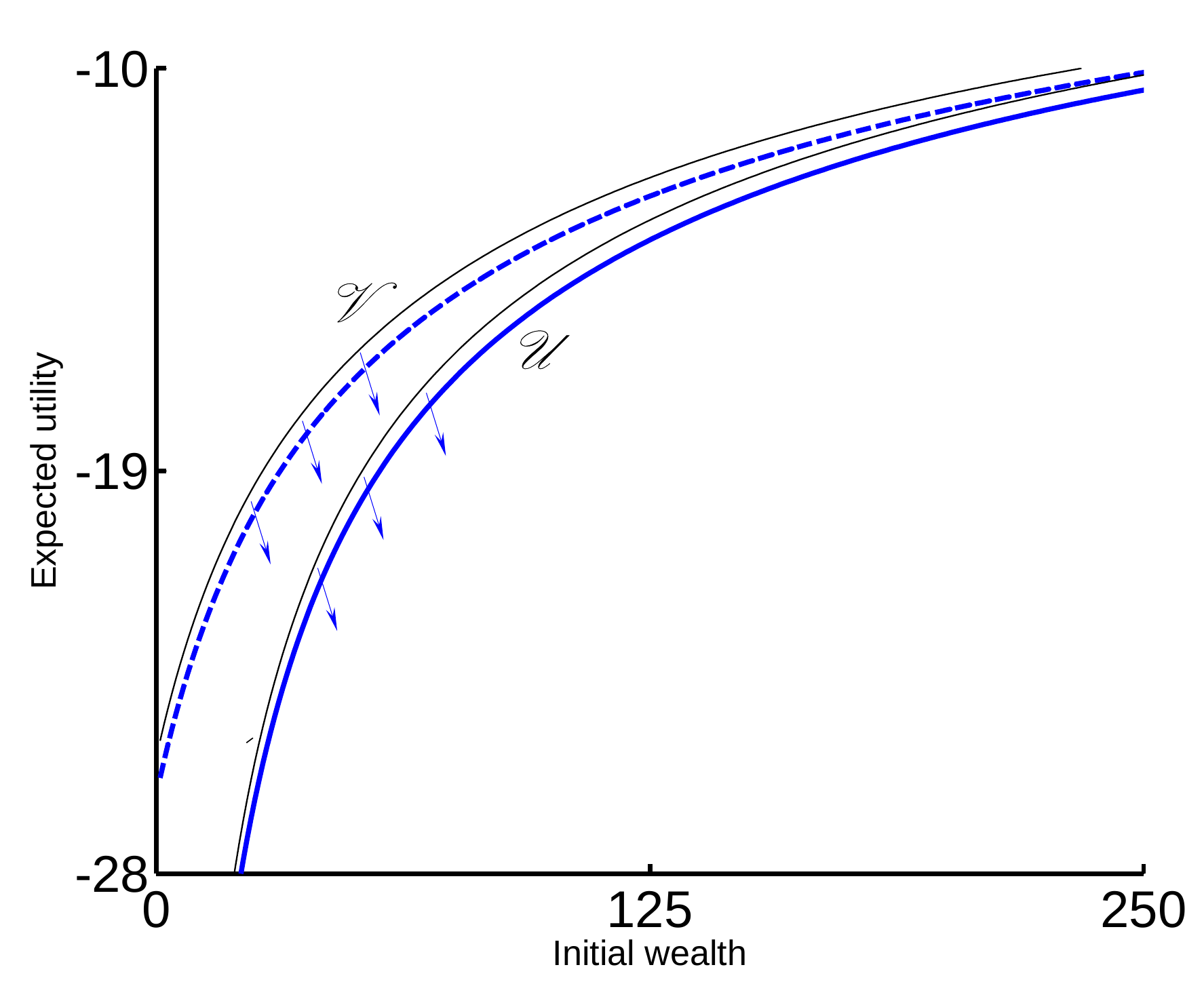}}\hspace{0pt}
  \subfigure[Density of death.]{\label{fig:StochMortality_Density}\includegraphics[scale=0.33]{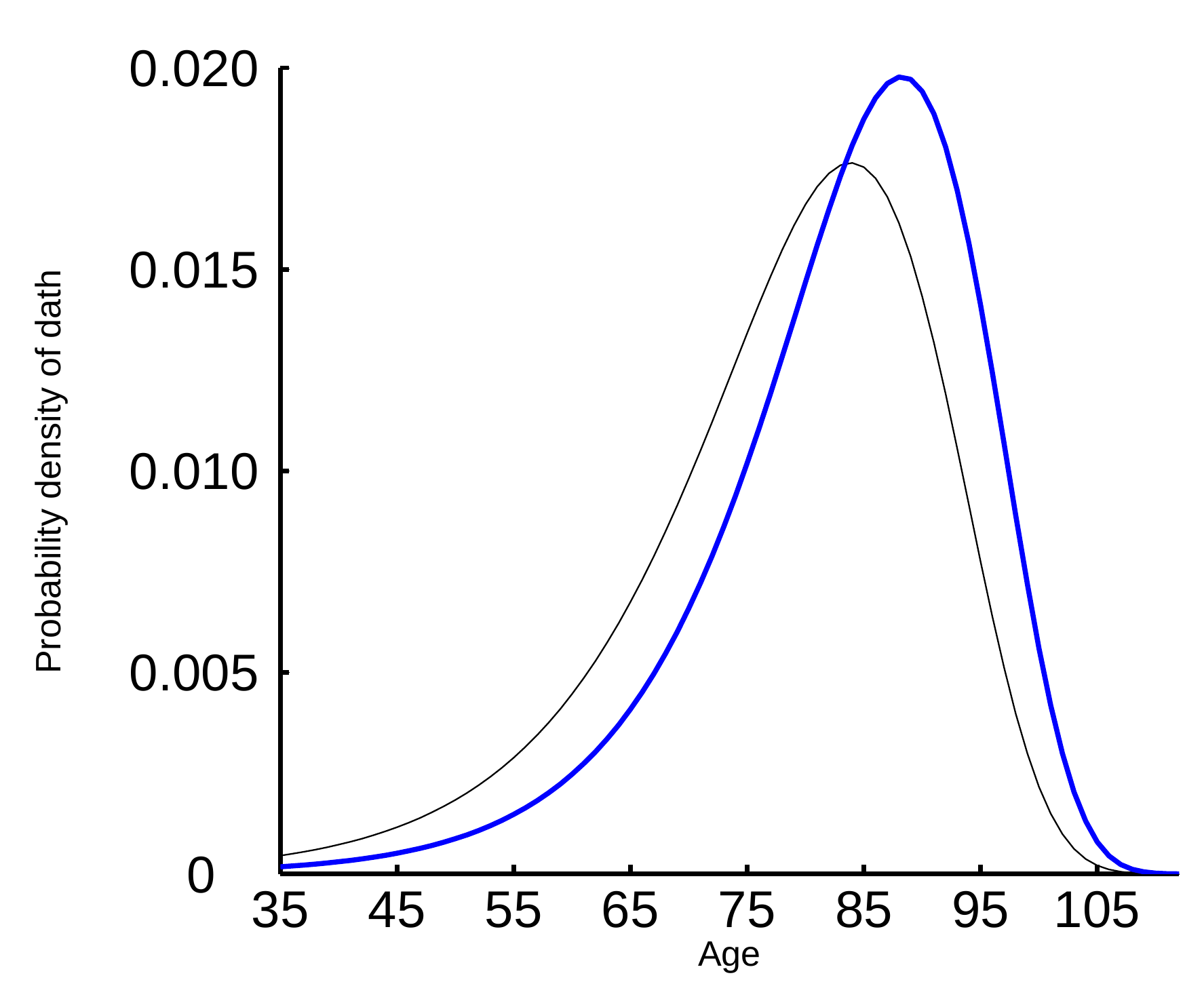}}
  \end{center}
\end{figure}

Through our analysis we consider a deterministic process for the force of mortality. Under this assumption, the indifference valuation -- in line with the previous literature -- depends on the difference between the interest rate $r$ and the guaranteed rate $h$. However it is interesting to simulate the effect arising from different mortality rates, because even if the indifference value (at time $t_0$) is still given by $L^*_0$, the value functions $\cal U$ and $\cal V$ change. Figure \ref{fig:StochMortality} show the effect of assuming different mortality specifications.

For instance, we compare the optimal expecter reward considering the same policyholder under a different subjective assessment of the  survival probability -- notice that in both cases we assume deterministic process for $\psg{}{}$, considering different scenarios. To make things easy, we keep referring to Table \ref{tb:GompertzEstimations}, comparing the estimations available in 1970 and in 2004. The same value functions plotted in figure \ref{fig:ValFunct_calUV_a} are compared with the ones calculated using data available in 2004.

In other words, if the policyholder could use a more optimistic  assessment for the survival probability, the convenience to by the option remains the same (i.e. $\cal V>\cal U$) but the expected utility is affected in the change in the value of $\phi$ and the negative value of $1-\gamma$.  The coefficient $\gamma$ expresses the policyholder's risk aversion over a larger trading horizon. For instance, the gap between the \textv{new} $\cal V$ and the \textv{old} $\cal V$ (as well as the \textv{new} $\cal U$ and the \textv{old} $\cal U$) reduces for smaller values of $\gamma$.
Eventually, for $\gamma \appl 0$, the policyholder does not suffer any impact from different mortality scenarios. Notice, however, that this claim is true because different scenarios do not change during the trading period, that is, there is no stochasticity other than the perturbation considered just at the initial time $t_0$.

%=================================================
 \section{Conclusions}
        \label{sec:Conclusions}
%=================================================

We value the guaranteed annuity option using an equivalent utility approach. The valuation is made from the policyholder's point of view. In a setting where interest rates are constant, we find  an explicit solution for the indifference problem, where power--law utility of consumption is assumed. In this setting we compare two strategies when the policy matures, and two strategies at the initial time. For the former we assume that, if the annuitant does not exercise the option, she first withdraws her  accumulated funds and then she seeks to solve a standard Merton problem under an infinite time horizon case. At the time when the policy matures, we compare the policyholder's expected reward associated to a policy embedding a guaranteed annuity option, and the one which arise from a policy that does not embed such an option. We find that the option's indifference price depends on the difference between the market interest rate $r$ and the guaranteed conversion rate $h$. Numerical experiments reveals that in periods characterized by high market interest rates, the value of the \gao turns out to be very small.  Finally, we also consider an auxiliary (and independent) problem in which we  compare the pure premium asked by the insurance company (for accumulating the funds up to the time of conversion) and the indifference price for the embedded option.

For future research, the present model can be generalized in several ways. First, the policyholder can be allowed to annuitize her wealth more than once during her retirement period. This fact leads us to consider an unrestricted market where the policyholder can annuitize anything at anytime, as defined by Milevsky \& Young \cite{MilevskyYoung_2007_ART_AnnuitizationAssetAllocation}.
Second, the financial market can be modeled considering a richer setting: stochastic interest rates and stochastic labor income. To this end, we recall the work of Koo \cite{Koo_ART_ConsumptionAndPortfolioSelection}. Third, and most important, in the present framework, the longevity risk is considered by comparing different scenarios, given by the survival tables available in 1970 and in 2004. For this, a more general stochastic approach, as proposed by Dahl \cite{Dahl_ART_StochasticMortality}, can be considered instead.

%********************************************************************
%********************************************************************

%------------------------------ BIBLIOGRAFIA --------------------------------

%\small

%GATHER{Bibliografia.bib}
%\bibliographystyle{plain}
%\bibliography{Bibliografia}

\begin{thebibliography}{10}

\bibitem{Bacinello_ART_FairValuationLifeInsuranceContractsWithEmbeddedOptions}
Anna~Rita Bacinello.
\newblock Fair valuation of life insurance contracts with embedded options.
\newblock {\em Association Suisse des Actuaries -- Assembl{\'e}e
  g{\'e}n{\'e}rale 2006}, 2006.

\bibitem{Ballotta_Haberman_ART_ValuationGAOs}
Laura Ballotta and Steven Haberman.
\newblock Valuation of guaranteed annuity convesion options.
\newblock {\em Insurance: Mathematics and Economics}, 33:87--108, 2003.

\bibitem{Ballotta_Haberman_ART_}
Laura Ballotta and Steven Haberman.
\newblock The fair valuation problem of guaranteed annuity options: The
  stochastic mortality environment case.
\newblock {\em Insurance: Mathematics and Economics}, 38:195--214, 2006.

\bibitem{Biffis_Millossovich_ART}
Enrico Biffis and Pietro Millossovich.
\newblock The fair value of guaranteed annuity options.
\newblock {\em Scandinavian Actuarial Journal}, (1):23--41, 2006.

\bibitem{Blake_Cairns_Dowd_ART_AnnuitizationOptions}
David Blake, Andrew Cairns, and Kevin Dowd.
\newblock Annuitization options.
\newblock {\em Prepared for the Global Issues in Insurance Regulation
  conference}, London, April 2002 2002.

\bibitem{Bolton_ART_ReservingForAnnuityGuarantees}
M.~J. Bolton and Collis P. A. George C.M. Knowles V. P. Whitehouse A.~J. Carr,
  D.~H.
\newblock Reserving for annuity guarantees.
\newblock {\em The Report of the Anuuity Guarantees Working Party}, November
  1997.

\bibitem{BoyleHardy_WP_GuarnAnnOpts}
Phelim Boyle and Mary Hardy.
\newblock Guaranteed annuity options.
\newblock {\em Working Paper -- University of Waterloo, Ontario, Canada N2L
  3G1}, 2002.

\bibitem{BoyleHardy_ART_GuaranAnnOpts}
Phelim Boyle and Mary Hardy.
\newblock Guaranteed annuity options.
\newblock {\em ASTIN Bulletin}, 33(2):125--152, 2003.

\bibitem{CairnsBlakeDowd_ART_ATwoFactorModel}
Andrew Cairns, David Blake, and Kevin Dowd.
\newblock A two-factor model for stochastic mortality with parameter
  uncertainty.
\newblock {\em CRIS Discussion Paper Series}, VI, 2005.

\bibitem{CairnsBlakeDowd_ART_PricingDeath}
Andrew Cairns, David Blake, and Kevin Dowd.
\newblock Pricing death: Frameworks for the valuation and securitization of
  mortality risk.
\newblock {\em CRIS Discussion Paper Series}, IV, 2006.

\bibitem{Carmona_BOOK_IndiffPricing}
Ren{\'e} Carmona, editor.
\newblock {\em Indifference Pricing -- Theory and Applications}.
\newblock Princeton University Press, 2009.

\bibitem{CARRIERE94_InvestigationOfGompertzLaw}
Jacques~F. Carriere.
\newblock An investigation of the gompertz law of mortality.
\newblock {\em Actuarial Research Clearing House}, 2:161--177, 1994.

\bibitem{Carriere94_UltimateParametricModel}
Jacques~F. Carriere.
\newblock A select and ultimate parametric model.
\newblock {\em Transactions of the Society of Actuaries}, 46(75-97), 1994.

\bibitem{Dahl_ART_StochasticMortality}
Mikkel Dahl.
\newblock Stocahstic mortality in life insurance: market reserves and
  mortality-linked insurance contracts.
\newblock {\em Insurance: Mathematics and Economics}, 35:113--136, 2004.

\bibitem{FreesCarriereValdez_ART_AnnuityValuation}
Edward~W. Frees, Jacques Carriere, and Emiliano Valdez.
\newblock Annuity valuation with dependent mortality.
\newblock {\em The Journal of Risk and Insurance}, 63(2):229--261, 1996.

\bibitem{GatzertKing_ART_AnalysisParticipatingLifeInsuranceContracts}
Nadine Gatzert and King Alexander.
\newblock Analysis of participating life insurance contracts: A unification
  approach.
\newblock {\em The Journal of Risk and Insurance}, 74(3):547--570, 2007.

\bibitem{Gerber_LIB}
Hans~U. Gerber.
\newblock {\em Life insurance mathematics}.
\newblock Springer, 2nd expanded, with exercises contributed by samuel h. cox
  edition, 1995.

\bibitem{GrasselliSilla_BuyerGAO}
Matheus~R. Grasselli and Sebastiano Silla.
\newblock A policyholder's indifference valuation for the guaranteed annuity
  option.
\newblock {\em Quaderni di Dipartimento, n. 27, November. \normalfont
  Dipartimento di Scienze Sociali, Facolt{\`a} di Economia \textv{Giorgio
  Fu{\`a}}, Universit{\`a} Politecnica delle Marche, Ancona, Italy.}, 2008.

\bibitem{Hardy_BOOK_InvestmentGuarantees}
Mary Hardy.
\newblock {\em Investment Guarantees -- Modeling and Risk Management for
  Equity-Linked Life Insurance}.
\newblock Wiley Finance. Wiley, 2003.

\bibitem{HodgesNeuberger}
Stewart~D. Hodges and Anthony Neuberger.
\newblock Optimal replication of contingent claims under transaction costs.
\newblock {\em Review Futures Markets}, 8:222--239, 1989.

\bibitem{KimOmberg_ART_DynamicNonmyopic}
Tong~Suk Kim and Edward Omberg.
\newblock Dynamic nonmyopic portfolio behavior.
\newblock {\em The Review of Financial Studies}, 9(1):141--161, Spring 1996.

\bibitem{Koo_ART_ConsumptionAndPortfolioSelection}
Hyeng~Keun Koo.
\newblock Consumption and portfolio selection with labor income: A continuous
  time approach.
\newblock {\em Mathematica Finance}, 8(1):49--65, January 1998.

\bibitem{Lee_ART_TheLeeCarterMethod}
Ronald Lee.
\newblock The lee-carter method for forecasting mortality, with various
  extensions and applications.
\newblock {\em North American Actuarial Journal}, 4(1):80--93, 2000.

\bibitem{Macdonald_ART_InternationalComparisonMortalityTrends}
A.~S. Macdonald, A.~J.~G. Cairns, P.~L. Gwilt, and K.~A. Miller.
\newblock An international comparison of recent trends in population mortality.
\newblock {\em British Actuarial Journal}, 4(1):3--141, 1998.

\bibitem{MarceauGaillardetz}
Etienne Marceau and Gaillardetz Patrice.
\newblock On life insurance reserves in a stochastic mortality and interest
  rates environment.
\newblock {\em Insurance: Mathematics and Economics}, 25:261--280, 1999.

\bibitem{MaroccoPitacco_ART_LongevityRisk}
Patrizia Marocco and Ermanno Pitacco.
\newblock Longevity risk and life annuity reinsurance.
\newblock {\em Working Paper -- Department of Matematica Applicata alle Scienze
  Economiche Statistiche e Attuariali \textv{Bruno de Finetti}}, 8/1997, 1997.

\bibitem{Merton69}
Robert~C. Merton.
\newblock Lifetime portfolio selection under uncertainty: The continuous-time
  case.
\newblock {\em The Review of Economics and Statistics}, 51(3):247--257, August
  1969.

\bibitem{Merton71}
Robert~C. Merton.
\newblock Optimum consumption and protfolio rules in a continuous-time model.
\newblock {\em Journal of Economic Theory}, 3:373--413, 1971.

\bibitem{Merton90}
Robert~C. Merton.
\newblock {\em Continuous-Time Finance}.
\newblock Blackwell Publishing, revised ed. -- foreword by paul a. samuelson
  edition, 1992.

\bibitem{Milevsky_LIB}
Moshe~A. Milevsky.
\newblock {\em The Calculus of Retirement Income -- Financial Models for
  Pensions Annuities and Life Insurance}.
\newblock Cambridge University Press, New York, NY, 2006.

\bibitem{MilevskyMooreYoung2006_ART_AssAllocAnnuityPurchaseAndFinancialRuin}
Moshe~A. Milevsky, Kristen~S. Moore, and Virginia~R. Young.
\newblock Asset allocation and annuity-purchase strategies to minimize the
  probability of financial ruin.
\newblock {\em Matematical Finance}, 16(4):647--671, 2006.

\bibitem{Milevsky_Posner_2001}
Moshe~A. Milevsky and Steven~E. Posner.
\newblock The tinatic option: Valuation of the guaranteed minimum death benefit
  in variable annuities and mututal funds.
\newblock {\em The Journal of Risk and Insurance}, 68(1):93--128, 2001.

\bibitem{Milevsky_Promislow_2001}
Moshe~A. Milevsky and S.~David Promislow.
\newblock Mortality derivatives and the option to annuitise.
\newblock {\em Insurance: Mathematics and Economics}, 29:299--318, 2001.

\bibitem{Milevsky_Promislow_Young_2005}
Moshe~A. Milevsky, S.~David Promislow, and Virginia~R. Young.
\newblock Financial valuation of mortality risk via the instantaneous sharpe
  ratio.
\newblock {\em IFID Center, Toronto, Canada}, 2005.

\bibitem{Milevsky_Promislow_Young_2006}
Moshe~A. Milevsky, S.~David Promislow, and Virginia~R. Young.
\newblock Killing the law of large numbers: Mortality risk premiums and the
  sharpe ratio.
\newblock {\em The Journal of Risk and Insurance}, 73(4):673--686, 2006.

\bibitem{MilevskyYoung2002_ART_OptimalAssetAllocation}
Moshe~A. Milevsky and Virginia~R. Young.
\newblock Optimal asset allocation and the real option to delay annuitization:
  It's not now-or-never.
\newblock {\em Preprint}, April, 13 2002.

\bibitem{MilevskyYoung_2003_ART_AnnuitizationAssetAllocation}
Moshe~A. Milevsky and Virginia~R. Young.
\newblock Annuitization and asset allocation.
\newblock {\em Preprint}, 2003.

\bibitem{MilevskyYoung2003_PRT_OptimalAnnuityPurchasing}
Moshe~A. Milevsky and Virginia~R. Young.
\newblock Optimal annuity purchasing.
\newblock {\em Preprint}, January 2003.
\newblock Downloadble at http://www.ifid.ca/research.htm.

\bibitem{MilevskyYoung_2007_ART_AnnuitizationAssetAllocation}
Moshe~A. Milevsky and Virginia~R. Young.
\newblock Annuitization and asset allocation.
\newblock {\em Journal of Economic Dynamic \& Control}, 31:3138--3177, 2007.

\bibitem{Milevsky98_ART_OptimalAssetAllocation}
Moshe~Arye Milevsky.
\newblock Optimal asset allocation towards the end of the life cycle: To
  annuitize or not to annuitize?
\newblock {\em The Journal of Risk and Insurance}, 65(3):401--426, 1998.

\bibitem{Milevsky2001}
Moshe~Arye Milevsky.
\newblock Optimal annuitization policies: Analysis of the options.
\newblock {\em North American Actuarial Journal}, 5(1):57--69, 2001.

\bibitem{MusielaZariphopoulou_ART_AnExampleOfIndifferencePrices}
Marek Musiela and Thaleia Zariphopoulou.
\newblock An example of indifference prices under exponential preferences.
\newblock {\em Finance and Stochastics}, 8:229--239, 2004.

\bibitem{OBrien_ART_GAOs}
Chris O'Brien.
\newblock Guaranteed annuity options: Five issues for resolution.
\newblock {\em CRIS Discussion Paper Series}, 8, 2002.

\bibitem{Olivieri_ART_UncertaintyInMortalityProgections}
Annamaria Olivieri.
\newblock Uncertainty in mortality projections: an actuarial perspective.
\newblock {\em Insurance: Mathematics and Economics}, 29:231--245, 2001.

\bibitem{OlivieriPitacco_ART_RenditeVitalizeLongevityRisk}
Annamaria Olivieri and Pitacco Ermanno.
\newblock Rendite vitalizie: longevity risk, garanzie demografiche, prolifi
  attuariali.
\newblock {\em Working Paper n. 22, CERAP Universit{\`a} Bocconi}, 2001.

\bibitem{OlivieriPitacco_ART_AnnuitisationGuaranteeAndUncertainty}
Annamaria Olivieri and Pitacco Ermanno.
\newblock Annuitisation guarantee and uncertainty in mortality trends.
\newblock {\em Working Paper n. 30, CERAP Universit{\`a} Bocconi}, 2003.

\bibitem{OlivieriPitacco_ART_ForecastingMortality}
Annamaria Olivieri and Ermanno Pitacco.
\newblock Forecasting mortality: An introduction.
\newblock {\em Working Paper n. 35, CERAP Univerit{\`a} Bocconi}, 2005.

\bibitem{Pelsser_ART_PricingAndHedgingGAOS}
Antoon Pelsser.
\newblock Pricing and hedging guaranteed annuity options via static option
  replication.
\newblock {\em Insurance: Mathematics and Economics}, 33:283--296, 2003.

\bibitem{Pelsser_PRT_PricingHedgingGuaranteedAnnuityOptions}
Antoon Pelsser.
\newblock Pricing and hedging guaranteed annuity options via static option
  replication.
\newblock {\em Preprint}, 2003.

\bibitem{Pitacco___ART_LongevityRiskInLivingBenefits}
Ermanno Pitacco.
\newblock Longevity risk in living benefits.
\newblock {\em CeRP -- Center for Research on Pensions and welfare Policies},
  2002.

\bibitem{Pitacco_ART_SurvivalModels}
Ermanno Pitacco.
\newblock Survival models in a dynamic context: a survey.
\newblock {\em Insurance: Mathematics and Economics}, 35:279--298, 2004.

\bibitem{Ruttermann_ART_MortalityTrendsWorldwide}
Markus R\"{u}ttermann.
\newblock Mortality trends worldwide.
\newblock {\em Risk Insights}, 3(4):18--21, 1999.

\bibitem{Schrager_ART_AffineStochasticMortality}
Schrager.
\newblock Affine stochastic mortality.
\newblock {\em Insurance: Mathematics and Economics}, 38:81--97, 2006.

\bibitem{Sorensen_ART_DynamicAssetAllocation}
Carsten S{\o}rensen.
\newblock Dynamic asset allocation and fixed income management.
\newblock {\em Journal of Financial and Quantitative Analysis}, 34(4):513--531,
  1999.

\bibitem{Stallard__Demographic}
Eric Stallard.
\newblock Demographic issues in longevity risk analysis.
\newblock {\em The Journal of Risk and Insurance}, 73(4):575--609, 2006.

\bibitem{Trigeorgis_ART_TheNatureOptionInteraction}
Lenos Trigeorgis.
\newblock The nature of option interactions and the valuation of investments
  with multiple real options.
\newblock {\em Journal of Financial and Quantitative Analysis}, 28(1):1--20,
  1993.

\bibitem{Wachter_ART_PortfolioConsumptionDecision}
Jessica~A. Wachter.
\newblock Portfolio and consumption decision under mean-reversting returns: An
  exact solution for complete markets.
\newblock {\em Journal of Financial and Quantitative Analysis}, 37(1):63--91,
  2002.

\bibitem{Wilkie_ART_ReservingPricingHedgingGAOs}
A.~D. Wilkie, H.~R. Waters, and Yang S.
\newblock Reserving, pricing and hedging for policies with guaranteed annuity
  options.
\newblock {\em British Actuarial Journal}, 9(2):263--425, 2003.

\bibitem{Willets_ART_MortalityEmprovement}
Richard Willets.
\newblock Mortality improvement in the united kingdom.
\newblock {\em Risk Insights}, 3(4):1--18, 1999.

\bibitem{Willets_ART_LongevityIn21Century}
Richard Willets, A.~P. Gallo, P.~A. Leandro, J.~L.~C. Lu, A.~S. Macdonald,
  K.~A. Miller, S.~J. Richards, N.~Robjohns, J.~P. Ryan, and H.~R. Waters.
\newblock Longevity in the 21st century.
\newblock {\em British Actuarial Journal}, 10(2):685--832, 2004.

\bibitem{Young03}
Virginia~R. Young.
\newblock Equity-indexed life insurance: Pricing and reserving using the
  principle of equivalent utility.
\newblock {\em North American Actuarial Journal}, 7(1):68--86, Jan. 2003.

\bibitem{Young_Zari_2002}
Virginia~R. Young and Thaleia Zariphopoulou.
\newblock Pricing dynamic insurance risks using the principle of equivalent
  utility.
\newblock {\em Scandinavian Actuarial Journal}, 4:246--279, 2002.

\bibitem{Zariphopoulou_BOOK_StochasticControlMethods}
Thaleia Zariphopoulou.
\newblock {\em Handbook of Stochastic Analysis and Applications}, chapter 12,
  \textv{Stochastic Control Methods in Asset Pricing}, pages 679--753.
\newblock Marcel Dekker, 2002.

\end{thebibliography}
%
%
%
\end{document}